\documentclass{elsart5}

\usepackage{graphicx}
\usepackage{amssymb}

\newcommand{\tal}{TlCuCl$_3$}

\begin{document}

\begin{frontmatter}

\title{Thermodynamics of the coupled spin-dimer system \tal\
close to a quantum phase transition}

\author[Colonia]{T. Lorenz\corauthref{Lorenz}}
\ead{lorenz@ph2.uni-koeln.de}
\author[Colonia]{S. Stark}
\author[Colonia]{O. Heyer}
\author[Colonia]{N. Hollmann}
\author[Mos]{A. Vasiliev}
\author[oos]{A. Oosawa}
\author[tan]{H. Tanaka}

\address[Colonia]{II.\,Physikalisches Institut, Universit\"{a}t zu
K\"{o}ln,Z\"{u}lpicher Str. 77, 50937 K\"{o}ln, Germany}
\address[Mos]{Department of Low Temperature Physics, Moscow State
University, Moscow 119992, Russia}
\address[oos]{Advanced Science Research Center, Japan Atomic Energy Research
Institute, Tokai, Ibaraki 319-1195, Japan}
\address[tan]{Department of Physics, Tokyo Institute of Technology, Oh-okayama,
Meguro-ku, Tokyo 152-8551, Japan}

\corauth[Lorenz]{}

\begin{abstract}
We present thermal expansion $\alpha $, magnetostriction and
specific heat $C$ measurements of \tal , which shows a quantum
phase transition from a spin-gap phase to a N\'{e}el-ordered ground
state as a function of magnetic field around $H_{C0}\simeq 4.8$~T.
Using Ehrenfest's relation, we find huge pressure dependencies of
the spin gap for uniaxial as well as for hydrostatic pressure.
For $T\rightarrow 0$ and $H\simeq H_{C0}$ we observe a diverging
Gr\"{u}neisen parameter $\Gamma(T)=\alpha / C$, in qualitative
agreement with theoretical predictions. However, the predicted
individual temperature dependencies $\alpha(T)$ and $C(T)$ are
not reproduced by our experimental data.

\end{abstract}


\end{frontmatter}

\section{Introduction}\label{}

Low-dimensional quantum magnets show very rich and fascinating
physical properties~\cite{lemmens03a}. As a starting point, one
may consider isolated spin-1/2 dimers with an antiferromagnetic
coupling $J$ causing a singlet ground state and an excited
triplet state, which are separated by an energy gap $\Delta=J$.
If such dimers are magnetically coupled to each other, a
multitude of different theoretical models can be constructed,
depending on the strength and the geometric arrangement of the
interdimer coupling(s) $J'$. As a consequence of one (or more)
non-zero $J'$, finite dispersion(s) of the triplet state evolve
along the respective direction(s) in reciprocal space, i.e., the
excited triplets may hop along different directions. Of
particular interest are one-dimensional (1D) chains with
alternating couplings $J$ and $J'$ between neighboring spins,
because of qualitatively different excitation spectra of the
alternating ($J'\neq J$) and the uniform ($J'=J$)
chain~\cite{uhrig96a}. Another example of 1D coupled spin dimers
is represented by so-called spin-ladders with the couplings
$J_{\perp}$ and $J_{\|}$ along the rungs and legs of the ladder,
respectively. The excitation gap of two-leg ladders is finite,
while it vanishes for a three-leg spin-ladder~\cite{dagotto96a}.
This difference is easily understood in the limit $J_{\perp}\gg
J_{\|}$, where the two-leg ladder can be viewed as weakly coupled
dimers and the three-leg ladder as an effective $S=1/2$ chain with
uniform chain coupling $J_{\|}$. With increasing number of legs,
the spin ladders approach the 2D antiferromagnetic square
lattice. Another well-studied system of 2D-coupled spin dimers is
the 2D Shastry-Sutherland model~\cite{shast81b}, which can be
generated from the 2D square lattice by introducing one
additional diagonal coupling $J_D$ on every second square. The
triangular arrangement of one $J_D$ and two $J$ causes a strong
frustration and for $J/J_D\lesssim 0.7$ the product state of
singlets on every diagonal is the exact ground state.

The above-mentioned models have been studied very intensively by
theoretical as well by experimental physicists during the last
decades, since a large number of materials became available which
rather well represent various types of these
models~\cite{lemmens03a}. Some cuprate examples are: the 2D square
lattice realized by the parent compound La$_2$CuO$_4$ of the
High-T$_c$'s, the spin-Peierls system CuGeO$_3$~\cite{hase93a},
spin-1/2 chain and ladder compounds, such as Sr$_2$CuO$_3$ and
SrCu$_2$O$_3$, respectively, as well as
Sr$_{14}$Cu$_{24}$O$_{41}$ containing both, spin chains and spin
ladders~\cite{dagotto99}, or the 2D Shastry-Sutherland model
which is realized in SrCu$_2$(BO$_3$)$_2$~\cite{kagey99}.

\begin{figure}[t]
\begin{center}
\includegraphics[width=.48\textwidth]{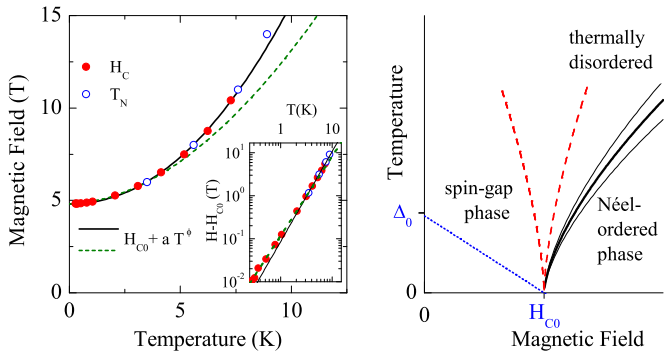}
\end{center}
\caption{Left: Phase boundary of \tal\ obtained from
magnetostriction ($\bullet$) and thermal expansion ($\circ$). The
dashed line is a fit of the form $H_C(T)=H_{C0}+a\cdot T^\phi$
for $T\le 2$~K, which yields $H_{C0}=4.82$~T and $\phi=1.85$,
while the solid line is a fit up to 8~K yielding $\phi=2.1$. The
$T$ dependence of $\phi$ is shown in the inset by the plot of
$H_C-H_{C0}$ versus $T$ on double-logarithmic scales, and is also
obtained by Quantum Monte Carlo calculations~\cite{nohadani2004}.
Right: Schematic view of the phase diagram around the quantum
phase transition from a spin-gap to a N\'{e}el-ordered ground state
at $H_{C0}$. The dotted line shows the spin-gap closing due to
the Zeeman splitting, and the dashed lines indicate the region of
enhanced quantum fluctuations at $T>0$. The thick solid line is
the phase boundary $T_N(H)$ and the thin solid lines indicate the
region of enhanced thermal fluctuations. } \label{phadi}
\end{figure}

\tal\ is another quantum spin system which has been intensively
studied in recent years~\cite{tanaka2001,Oosawa2002d}. From the
structural point of view this compound contains a ladder-like
arrangement of Cu$^{2+}$ ions~\cite{tanaka2001}. The main magnetic
coupling $J\simeq 5.5$~meV is present along the rungs, but there
are various additional, rather large couplings $J'$ present along
different other lattice
directions~\cite{Nikuni2000,matsumoto2002,ruegg03,matsumoto2004}.
Thus, the magnetic system of \tal\ should be viewed as a set of
three-dimensionally coupled spin dimers. The inter-dimer
couplings $J'$ cause a strong dispersion of the triplet
excitations, and as a consequence the minimum singlet-triplet
$\Delta_m\simeq 0.7$~meV is much smaller than $J$. A moderate
field of about 5~T is already sufficient to close $\Delta_m$ and
induces a N\'{e}el order with staggered magnetization perpendicular
to the applied field. If there is no magnetic anisotropy in the
plane perpendicular to the applied field, this transition is in
the same universality class as the Bose-Einstein condensation
(BEC) and it is often termed a BEC of magnons. In the
zero-temperature limit, it represents an example of a quantum
phase transition~\cite{vojta03}, whose control parameter is the
magnetic field strength (see Fig.~\ref{phadi}). In the vicinity
of a quantum critical point (QCP) anomalous temperature
dependencies are expected for various physical properties, as
e.g.\ specific heat $C$, susceptibility $\chi$, thermal expansion
$\alpha$, (and resistivity $\rho$ for metals)~\cite{stewart01a}.
In particular, a divergence of the so-called Gr\"{u}neisen parameter
$\Gamma = \alpha /C$ is expected, when a pressure-dependent QCP is
approached~\cite{zhu03a}. Experimentally, a diverging $\Gamma(T)$
has indeed been observed for different heavy-fermion
compounds~\cite{kuechler03a,kuechler04a}. Since the phase
transition of \tal\ is extremely sensitive to
pressure~\cite{Oosawa2003a,tanaka2003a,johannsen05a,goto05a} and
the control parameter may be easily tuned by a variation of the
field, this compound is ideally suited to study such generic
properties of a QCP.

We present high-resolution measurements of the uniaxial thermal
expansion $\alpha_i=\partial \ln L_i/\partial T$ and the
magnetostriction $\epsilon_i =[L_i(H)-L_i(0)]/L_i(0)$ along
different lattice directions $i$ ($L_i$ is the respective sample
length along $i$) as well as specific heat $C$ and magnetization
$M$ data. The length changes have been measured down to 250~mK by
a home-built capacitance dilatometer and $C$ by a home-built
calorimeter for $T \gtrsim 500$~mK, while the magnetization has
been studied by a commercial vibrating sample magnetometer
(Quantum Design) for $T \gtrsim 1.9$~K. All properties have been
studied in magnetic fields up to 14~T. Since \tal\ easily cleaves
along the (010) and (10$\bar{2}$) planes of the monoclinic
structure, we measured $L_i(H,T)$ perpendicular to these planes
on a single crystal of dimensions $1.7 \times 1.5$~mm$^2$
perpendicular to (010) and (10$\bar{2}$), respectively. In
addition, the [201] direction, which is perpendicular to both
other directions, was measured on a second crystal of length
$L_{[201]}=4.4$~mm. For all three measurement directions $i$ the
magnetic field was applied along the same direction, namely
perpendicular to the (10$\bar{2}$) plane.

\section{Results and Discussion}

\begin{figure}[t]
\begin{center}
\includegraphics[width=.48\textwidth]{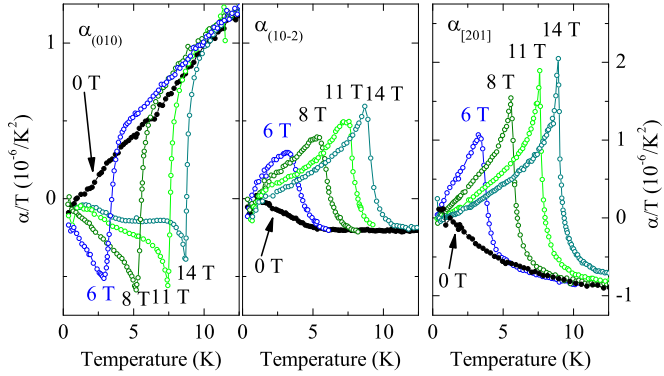}
\end{center}
\caption{Thermal expansion along different directions for various
magnetic fields applied perpendicular to the $(10\overline{2})$
cleavage plane of \tal . Note the different scale for the $[201]$
direction.} \label{alpha}
\end{figure}

\begin{figure}[t]
\begin{center}
\includegraphics[width=.48\textwidth]{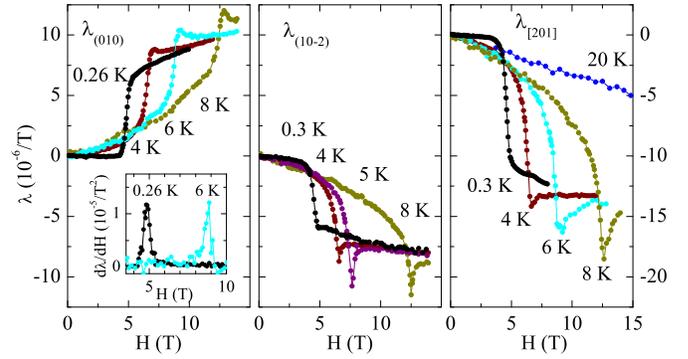}
\end{center}
\caption{Magnetostriction coefficient $\lambda_i \partial
\epsilon_i /\partial H$ measured along different directions $i$ at
various temperatures. In all cases the magnetic field has been
applied perpendicular to the $(10\overline{2})$ plane (see text).
} \label{ms}
\end{figure}

Fig.~\ref{alpha} shows $\alpha_i$ measured along all three
directions for different magnetic fields. In zero field,
$\alpha_i$ has no visible anomalies, but is strongly anisotropic.
For $H \gtrsim 5$~T, pronounced anomalies develop and shift
systematically towards higher $T$ with increasing $H$. The
$\alpha_i$ curves for $i=(010)$ and $(10\bar{2})$ agree well with
our previous results measured on a different crystal for $T
\gtrsim 3$~K~\cite{johannsen05a,johannsen06a}. The anomalies of
$\alpha_i$ signal large uniaxial pressure dependencies of $T_N$,
which are related to $\alpha$ and $C$ via Ehrenfest's relations
\begin{equation}
\frac{\partial T_N}{\partial p_i} = V_m T_N \frac{\Delta
\alpha_i}{\Delta C}
 \,\,\, \mbox{ and } \,\,\,
\frac{\partial H_C}{\partial p_i} = V_m \frac{\Delta
\lambda_i}{\Delta\chi } \,\, . \label{Ehr}
\end{equation}
Here, $V_m$ is the molar volume and $\Delta \alpha_i$ and $\Delta
C$ denote the respective mean-field jumps at $T_N$. The second
expression  of Eq.~(\ref{Ehr}) relates the pressure dependencies
of the transition field to the jumps of $\lambda_i=\partial
\epsilon_i/\partial H $ and of the differential magnetic
susceptibility $\chi=\partial M / \partial H $. For $i=(010)$ and
$(10\bar{2})$, the $\partial T_N/\partial p_i$ largely cancel
each other under hydrostatic pressure, since the anomalies are of
comparable magnitudes but of opposite signs. Thus, the
hydrostatic pressure dependence $\partial T_N/\partial p_{hydro}$
is essentially determined by the sign and size of the anomaly of
$\alpha_{201}$. Obviously, the anomalies of $\alpha_{201}$ are
the largest ones and their positive signs mean that $T_N$
drastically increases for uniaxial pressure along $[201]$ as well
as for hydrostatic pressure. Fig.~\ref{ms} shows the field
derivatives of the magnetostriction. Again, we find the
characteristic anisotropy that the anomalies of $\lambda_{010}$
and $\lambda_{10\bar{2}}$ are of similar sizes but opposite
signs, while significantly larger anomalies are present for
$\lambda_{201}$. Thus, the hydrostatic pressure dependence of
$H_C$ is again essentially identical to that for uniaxial
pressure along $[201]$.

In Ref.~\cite{johannsen05a} we have shown that the
pressure-dependencies of $T_N$, $H_C$, and the magnetic
susceptibility in the paramagnetic phase can be traced back to the
pressure-dependent changes of a single parameter, which in the
case of \tal\ is the intra-dimer coupling $J$. This conclusion
was based on the measurements perpendicular to the $(010)$ and
$(10\bar{2})$ planes, and is fully confirmed by the additional
new data along the $[201]$ direction. From the zero-temperature
extrapolations of the uniaxial pressure dependencies of $H_C$ for
$i=(010)$ and $(10\bar{2})$ we estimated $\partial \ln \Delta_m
/\partial p_i \simeq + 190$~\%/GPa and $\simeq -180$~\%/GPa.
Since the anomalies for the $[201]$ direction are about twice as
large, we obtain $\partial \ln \Delta_m /\partial p_i \simeq
-370$~\%/GPa for pressure along $[201]$ and $\simeq -360$~\%/GPa
for hydrostatic pressure. The latter value is in reasonable
agreement with direct measurements under hydrostatic pressure,
which yield $\partial \ln \Delta_m /\partial p_{hydro} \simeq
-400$~\%/GPa for the initial slope at ambient
pressure~\cite{goto05a}.

The shape of the $\alpha_i$ anomalies is typical for a
second-order phase transition with a pronounced mean-field
contribution causing a jump $\Delta \alpha $ at $T_N$,
superimposed by fluctuations causing a divergence $\alpha\propto
t^\nu$ with the reduced temperature $t=|T-T_N|/T_N$ and the
critical exponents $\nu$ depending on the universality class of
the phase transition. On approaching $H_{C0}=H_C(T\rightarrow
0)\simeq 4.8$~T, the $\alpha_i$ anomalies broaden to some extent
(see below). The $\lambda_i$ anomalies also show a pronounced
fluctuation contribution for $T
> 2$~K, but become more jump-like for lower $T$. The changing
shapes of both, the $\lambda_i$ and the $\alpha_i$ anomalies can
be intuitively understood from Fig.~\ref{phadi}, because (i) the
absolute temperature region around the phase boundary where
fluctuations become important decreases with decreasing $T_N$,
and (ii) close to $H_{C0}$ the phase boundary is crossed with a
very small slope as a function of $T$, so that the fact that
$H_C$ is not infinitely sharp becomes more and more important. As
a criterion for $H_C$, we used the maximum of the second
derivatives $\partial^2 \epsilon_i/\partial H^2$, whose full
widths at half maximum amount to $\simeq 0.45$~T (see Inset of
Fig.~\ref{ms}). We suspect that this width mainly arises from
internal stresses, which broaden the transition due to the strong
(uniaxial) pressure dependencies of $H_C$. We have also
investigated, whether there is a finite hysteresis of $H_C$ by
measuring $\epsilon_i(H)$ with increasing and decreasing $H$. For
a drift rate of $\pm 0.1$~T/min we obtain a difference
$H_C^{up}-H_C^{down}\simeq 0.04$~T, which does hardly change with
temperature. At $T=0.3$~K we also measured with $\pm 0.01$~T/min
and found $H_C^{up}-H_C^{down}\simeq 0.01$~T, i.e. , the observed
hysteresis partly arises from the finite field drift. Thus, we
regard the phase transition as a second-order one with a weak
first-order contribution. The first-order contribution most
probably arises from the large spin-lattice coupling, which may
drive a second-order into a first-order
transition~\cite{murata77}. It is also possible that a transition
transforms from second to first order, when $T_c$ is suppressed
towards 0~K by an external parameter. However, the weak
temperature-dependence of the hysteresis observed in \tal\ does
not give any evidence for such a scenario, and we suspect that
the transition of \tal\ remains (almost) continuous down to
lowest $T$.

As mentioned above, a diverging Gr\"{u}neisen parameter $\Gamma(T
\rightarrow 0)$ has been predicted at $H=H_{C0}$~\cite{zhu03a}.
Before comparing our experimental data to this prediction, we will
discuss from a phenomenological point of view what may be
expected for $\Gamma$ when the QCP is approached along different
routes in the phase diagram. Using Maxwell's relations one finds
that $\alpha=-\partial S/\partial p$, while $C/T=\partial
S/\partial T$. If a thermodynamic system can be described by a
single energy scale $E$, its entropy $S$ only depends on the ratio
$T/E$, i.e.\ $S(T,E)=S(T/E)$ with a model-dependent function
$S(x)$. Comparing the $T$- and $p$-derivatives of $S(T/E(p))$
yields the Gr\"{u}neisen scaling
\begin{equation}
\Gamma=\frac{\alpha(T)}{C(T)}=\frac{\partial \ln E}{\partial p}
\,\, , \label{Gru}
\end{equation}
which is temperature independent. Prominent examples of such
systems are the Debye model with $E=\Theta_D$, the almost free
electron gas with $E=E_F$, or magnetically ordered states with
exchange coupling $J=E$ (for $T\ll T_c$). In real systems, one
usually observes a weakly $T$-dependent $\Gamma(T)$, which is due
to the fact that the above-mentioned single-parameter models only
consider the leading energy scale and neglect others. If several
energy scales $E_i$ are equally important, the individual $E_i$
usually have different pressure dependencies, and therefore the
$p$- and $T$-derivatives are, in general, not (almost)
proportional to each other. However, it is possible that for
different temperature regions different $E_i$'s are dominant and
in the respective regions $\Gamma\simeq \partial \ln E_i/\partial
p$ holds. An example is a coupled spin-dimer system as it is
realized in \tal . For high temperatures ($T\gg J,J'$), the
behavior is determined by the average spin gap $\langle \Delta
\rangle$, while the minimum gap $\Delta_{m}$ becomes dominant at
$T\ll \Delta_{m}$. Thus, $\Gamma(T)$ varies from $\simeq \partial
\ln \Delta_{m} /\partial p$ to $\simeq \partial \ln \langle
\Delta \rangle /\partial p$  with increasing $T$. In \tal , the
zero-field gap $\Delta_{m}^0\simeq 8$~K linearly decreases with
$H$~\cite{ruegg03}, i.e.\ $\Delta_{m}(H)=\Delta_{m}^0-h$ where
$h=g\,H/k_B$, $g$ is the $g$-factor and $k_B$ Boltzmann's
constant. Thus, at a given field $H<H_{C0}$ the temperature,
below which $\Gamma$ is expected to approach a constant,
decreases with $H$. For $H\rightarrow H_{C0}$, however,
$\Gamma(T\ll\Delta_{m})\simeq \frac{1}{\Delta_{m}(H)}
\frac{\partial \Delta_{m}^0}{\partial p}$ diverges, because
$\Delta_{m}\rightarrow 0$.

\begin{figure}[t]
\begin{center}
\includegraphics[width=.48\textwidth]{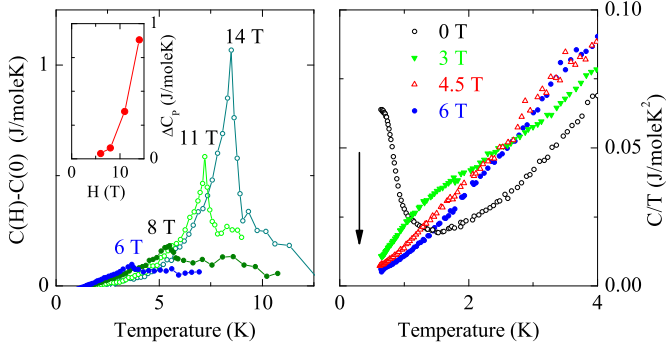}
\end{center}
\caption{Left: Difference of the specific heat $C$ in finite and
zero magnetic field. The inset shows the anomaly heights as a
function of field. Right: Expanded view of the low-temperature
$C/T $ for different fields below and above the critical field
$H_{C0}\simeq 4.8$~T. The arrow indicates increasing
magnetic-field strength. (see text)}\label{cp}
\end{figure}

For $H>H_{C0}$, i.e.\ in the ordered phase, the characteristic
low-temperature energy scale is given by the spin wave velocity
$v$. To our knowledge, the exact dependence of $v$ on $H-H_{C0}$
is not known, but it is quite natural that $v$ disappears when
$H_{C0}$ is approached. For simplicity, we assume $v \propto
\left( H-H_{C0}\right)^n$ with $n>0$ and derive
$\Gamma(T\rightarrow 0) \simeq
\partial \ln v /\partial p \propto \frac{1}{H-H_{C0}}
\frac{-\partial H_{C0}}{\partial p}$. Thus, the Gr\"{u}neisen
parameter is again expected to approach a constant, which diverges
for $H\rightarrow H_{C0}$, but the sign of the divergence for
$H>H_{C0}$ is opposite to that for $H<H_{C0}$.

In a next step we will approach the QCP along the phase boundary
$T_N(H)$ for $H>H_{C0}$. For clarity, we use the approximation
$T_N(H)=b\left(H-H_{C0}\right)^\varphi$ with $\varphi = 1/\phi
\rightarrow 2/3$ for $H \rightarrow H_{C0}$ (see Fig.~\ref{phadi}
and Refs.~\cite{nohadani2004,wessel2001}), and calculate $\partial
T_N/\partial p= -\varphi b\left(H-H_{C0}\right)^{(\varphi-1)}\cdot
\partial H_{C0}/\partial p$. Obviously, $\partial T_N/\partial p$
diverges for $H \rightarrow H_{C0}$ for $\varphi <1$, and the sign
of this divergence is opposite to the sign of $\partial
H_{C0}/\partial p$. We emphasize that this result follows from
the infinite slope of the phase boundary for $H\rightarrow
H_{C0}$, and does not depend on the particular choice of $T_N(H)$.
Since the pressure dependence of $T_N$ is given by Ehrenfest's
relation~(\ref{Ehr}), the ratio $\Delta \alpha / \Delta C$ of the
thermal-expansion and specific-heat anomalies at $T_N$ is
expected to diverge for $H\rightarrow H_{C0}$. From
Eq.~(\ref{Ehr}), it is also clear that the vanishing $T_N $ would
cause a divergence of $\Delta \alpha / \Delta C$ even if the
slope of the phase boundary was finite and $\partial T_N/
\partial p$ would thus not diverge for $H\rightarrow H_{C0}$.

Let us summarize the above considerations. On approaching $H_{C0}$
we expect (i) for $H<H_{C0}$ that $\Gamma(T\ll
\Delta(H))\rightarrow
\partial \ln \Delta(H) /\partial p$, which diverges for
$H\rightarrow H_{C0}$, (ii) for $H>H_{C0}$ a similar divergence
of $\Gamma(T\rightarrow 0)$, but of the opposite sign, and (iii)
a divergence of the ratio $\Delta \alpha/\Delta C$, which has the
same (opposite) sign as the divergence of $\Gamma$ above (below)
$H_{C0}$. Since the above considerations are rather general, one
may expect a divergence of $\Gamma(T)$ close to many kinds of
transitions, whose $T_c$ is suppressed to 0~K. To obtain more
information about a quantum phase transition, one has to consider
the actual temperature dependencies $\alpha(T)$, $C(T)$, and/or
$\Gamma(T)$, as it has been done e.g.\ by the authors of
Refs.\cite{zhu03a,garst05a,fischer05a}.

\begin{figure}[t]
\begin{center}
\includegraphics[width=.48\textwidth]{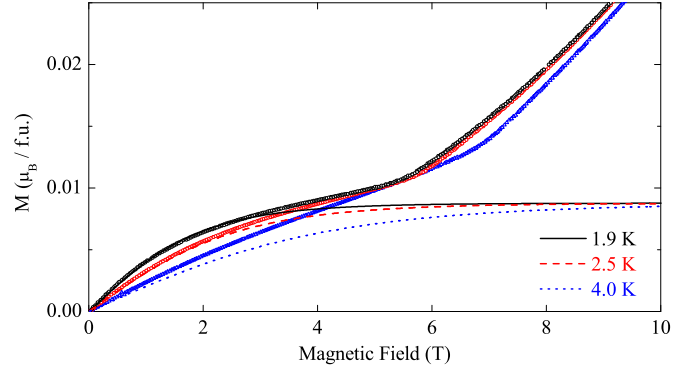}
\end{center}
\caption{Magnetization (symbols) of \tal\ at different
temperatures. The finite magnetization in the gap region can be
well reproduced by Brillouin functions (lines) assuming magnetic
impurities with spin $S=1$. (see text)} \label{mag}
\end{figure}

In Fig.~\ref{cp} we show the specific heat anomalies for
different magnetic fields. In agreement with
Ref.~\cite{Oosawa2001} we find rather small anomalies even for
the largest field, and their magnitude rapidly decreases when
$H_{C0}$ is approached (see Inset of Fig.~\ref{cp}). Since the
magnitude of the respective $\alpha_i$ anomalies changes much
less with field (see Fig.\ref{alpha}), the expected divergence of
$\partial T_N/\partial p_i$ for $H\rightarrow H_{C0}$ is obviously
confirmed by the experimental data, since the denominator in
Eq.~(\ref{Ehr}) vanishes. This is the case for all three
directions of uniaxial as well as for hydrostatic pressure. In
the right panel of Fig.~\ref{cp} we show an expanded view of the
low-temperature behavior of some $C(T,H)/T$ curves. For zero
field, the onset of an anomaly can be clearly
seen~\cite{remarkCp}, and this anomaly is suppressed above about
3~T. We suspect that this anomaly arises from an ordering of
magnetic impurities. The presence of such impurities is also
evident from the finite magnetization in the gap region $H <
H_{C0}$ at low $T$, which can be well reproduced by Brillouin
functions (see Fig.~\ref{mag}). The corresponding fits of the
data at $T=1.9$~K and 2.5~K yield 0.4\,\% of magnetic impurities
with spin $S=1$. The Brillouin function calculated for the same
parameters and $T=4$~K is somewhat smaller than the experimental
data for $H\gtrsim 1$~T. This is expected, because at this higher
temperature a sizeable magnetization from excited triplets is
already present. Probably, the $S=1$ impurities are mostly
ferromagnetically aligned spin dimers, because the intra-dimer
coupling between the spin-1/2 Cu$^{2+}$ ions arises from a
$\simeq 96^\circ$ superexchange via the $p$ orbitals of the
Cl$^-$ ions, which is very sensitive to changes in the bond
angle. The impurities strongly influence the low-temperature
behavior of $C(T)$ for low fields, but become much less
influential at higher fields because the moments are almost
completely saturated for $H\gtrsim 3$~T and $T\lesssim 2$~K. It
is, however, unclear to what extent the impurities may change the
(critical) behavior close to the phase transition.

\begin{figure}[t]
\begin{center}
\includegraphics[width=.48\textwidth]{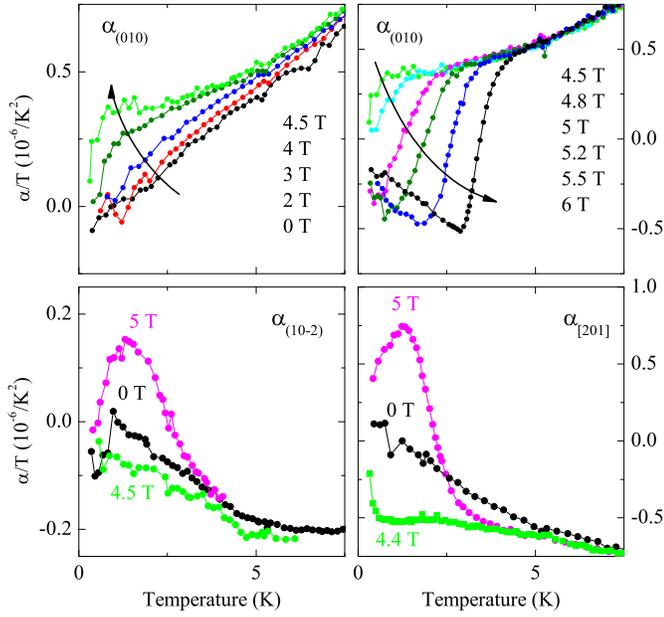}
\end{center}
\caption{Top: Thermal expansion perpendicular to the $(010)$ plane
for different magnetic fields below (left) and above (right) the
critical field. The arrows indicate increasing magnetic-field
strength. The bottom panels show some $\alpha_i/T$ curves for the
other directions.} \label{alpT}
\end{figure}

Fig.~\ref{alpT} shows an expanded view of the low-$T$ behavior of
$\alpha_i/T$. In zero field, $\alpha_i/T$ continuously approaches
zero for $T\rightarrow 0$, while for larger $H$ it shows a
pronounced shoulder, which systematically increases and reaches a
maximum slightly below $H_{C0}$. For larger fields clear
anomalies with a sign change of $\alpha_i/T$ occur, and these
anomalies systematically sharpen with further increasing field.
The behavior is essentially the same for all three directions,
only the magnitudes and signs are different~\cite{remarkalp}. As
already mentioned, we attribute the broadening of the anomalies
when $H_{C0}$ is approached from larger fields to the finite
width of the phase transition. This also explains that the
$\alpha_i(T)$ curves show anomalies already for $H \gtrsim
4.6$~T, i.e.\ below $H_{C0}\simeq 4.8$~T determined by the
magnetostriction measurements.

\begin{figure}[t]
\begin{center}
\includegraphics[width=.48\textwidth]{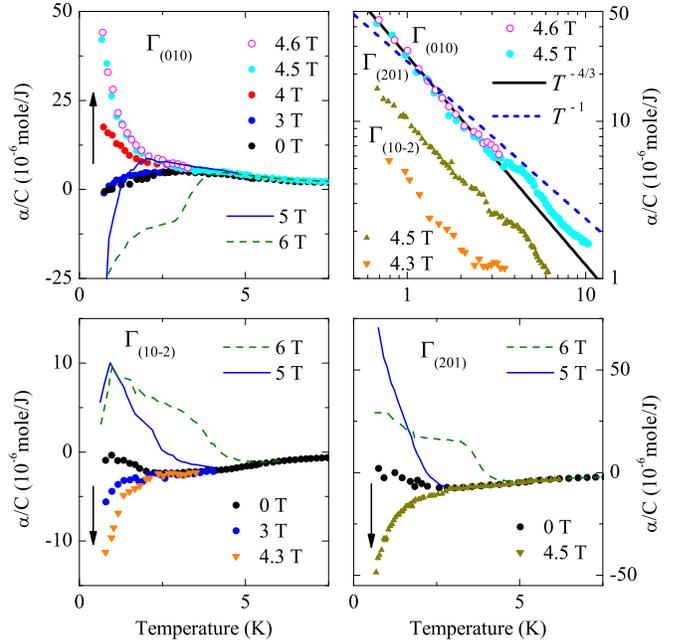}
\end{center}
\caption{Gr\"{u}neisen parameters $\Gamma_i$ obtained from the thermal
expansion along different measurement directions $i$ in magnetic
fields below (symbols, increasing field strength is indicated by
the arrows) and above (lines) the QCP at $H_{C0}\simeq 4.8$~T. The
upper right panel shows $\Gamma_i(T)$ for $H\lesssim H_{C0}$ on
double-logarithmic scales. Here, the negative $\Gamma_i(T)$ for
$i=(10\overline{2})$ and $[201]$ have been divided by a factor of
$-2$ and $-3$, respectively. The solid line is a power-law fit of
$\Gamma_{010}$ yielding $1/T^{4/3}$ and, for comparison, the
expected $1/T$ behavior is shown by the dashed line (see text).}
\label{Ga}
\end{figure}

In Fig.~\ref{Ga} we present  $\Gamma_i$ for different magnetic
fields. For all three directions $\Gamma_i$ shows the tendency to
diverge with decreasing $T$ for $H\simeq 4.5$~T. For lower $H$,
the $\Gamma_i(T)$ follow the same curve at higher $T$, but seem
to approach finite values for $T\rightarrow 0$. The magnitudes of
these limiting values increases with increasing field and the
temperature, below which the deviation sets in, decreases. The
$\Gamma_i(T)$ for $H>H_{C0}$ also follow the general curve at
higher $T$, until a large anomaly signals the crossing of the
phase boundary. For all three directions, the magnitudes of these
anomalies drastically increase with decreasing $H$ and the signs
are opposite to the respective signs of the diverging
$\Gamma_i(T)$ for $H< H_{C0}$.

On this qualitative level, our experimental data of $\Gamma_i(T)$
very well confirm the behavior, which one can expect from the
above considerations of the gap closing for $H<H_{C0}$ and the
shape of the phase boundary for $H>H_{C0}$. For a deeper
understanding, one has to compare the experimental data
quantitatively to theoretical predictions. According to
Ref.~\cite{zhu03a} the following temperature dependencies are
expected $H=H_{C0}$:
\begin{equation}
C/T \propto \sqrt{T} \,\, , \,\,\,\, \alpha/T \propto 1/\sqrt{T}
\,\, \mbox{ , and } \,\,\,\, \Gamma \propto 1/T \,\, .
\label{temps}
\end{equation}

In the upper right panel of Fig.~\ref{Ga} we show the diverging
$\Gamma_i(T)$ for $H\lesssim H_{C0}$ on double-logarithmic scales.
Because of the negative signs, $\Gamma_{10\overline{2}}$ and
$\Gamma_{201}$ have been divided by the factors $-2$ and $-3$,
respectively. Apparently, the slope is the same within
experimental accuracy. The solid line is a power-law fit of
$\Gamma_{010}$ which yields $T^{-4/3}$ and describes the
experimental data reasonably well for about one decade. For
comparison, the predicted $1/T$ behavior is also shown (dashed
line). In view of the fact that the theoretical prediction only
considers the irregular contributions of $C$ and $\alpha_i$,
while the experimental data also contain the phononic
contributions of $C$ and $\alpha_i$, one may tend to the
conclusion that our data nicely confirm the theoretical
prediction for $\Gamma(T)$. However, the agreement between theory
and experiment becomes much worse when the individual temperature
dependencies of $C/T$ and $\alpha_i/T$ are considered. Neither
the data of Fig.~\ref{cp} nor those of Fig.~\ref{alpT} give any
indication to follow the predicted temperature dependencies of
Eq.~(\ref{temps}). Concerning the specific heat data, one might
argue that the predicted $\sqrt{T}$ behavior is difficult to
identify because of the phononic contribution and the influence
of the magnetic impurities. This argument is less convincing for
$\alpha_i(T)/T$, since (i) the predicted divergence should be
seen despite a (regular) phononic contribution and (ii) the
ordering of the magnetic impurities does not cause a sizeable
anomaly in the zero-field data~\cite{remarkalp}. Thus, we
conclude that our present experimental data of \tal\ do not
confirm the theoretical predictions~\cite{zhu03a}. However, our
data do not disprove the theoretical predictions either.
Experimentally, one can suspect that in order to observe the
predicted temperature dependencies it would be necessary to study
(i) samples with significantly reduced transition widths and less
magnetic impurities, and (ii) it might be necessary to extend the
measurements to lower temperatures. Moreover, the theoretical
predictions have been calculated for clean and isotropic systems.
It is not clear to what extent the temperature dependencies of
$\alpha$ and $C$ are influenced by disorder or a finite magnetic
anisotropy. The latter is reflected in the $\simeq 10$\,\%
anisotropy of the $g$ factors of \tal\ for different magnetic
field orientations~\cite{glazkov04a}.

\section{Summary}

In summary, we have presented high-resolution measurements of
thermal expansion and magnetostriction along different lattice
directions of \tal . Both quantities show very pronounced and
strongly anisotropic anomalies at the phase boundary of the
field-induced N\'{e}el order $T_N(H)$ for $H > H_{C0}$, and signal
very large and strongly anisotropic uniaxial pressure dependencies
of the transition temperatures and fields. The hydrostatic
pressure dependence of the spin gap $\Delta_m$ for $H\rightarrow
H_{C0}$ obtained from our data using Ehrenfest's relations is in
reasonable agreement with the value observed by direct
measurements under hydrostatic pressure. In addition, our data
confirm the diverging pressure dependencies $\partial T_N
/\partial p_i $ for $H\rightarrow H_{C0}$, which are expected
from the infinite slope $\partial T_N /\partial H$ of the phase
boundary for $H\rightarrow H_{C0}$, i.e.\ when the quantum
critical point is approached. For $H< H_{C0}$, the Gr\"{u}neisen
parameters $\Gamma_i =\alpha_i/C$ are expected to approach
constant values for $T\ll \Delta(H)$, which diverge for
$H\rightarrow H_{C0}$, and $\Gamma_i(T) \propto 1/T$ has been
predicted at $H = H_{C0}$. In fact, the experimental $\Gamma_i(T)$
for all three measurement directions are in qualitative agreement
with these expectations. However, the temperature dependencies
predicted for the individual quantities $\alpha_i(T)$ and $C(T)$
are not observed experimentally. For $H\simeq H_{C0}$ the low-$T$
behavior of both $\alpha_i$ and $C$ is influenced by the finite
transition width and for lower fields at least $C(T)$ is also
affected by the presence of magnetic impurities. Thus, future
measurements on samples of improved quality as well as
calculations considering the influence of disorder and weak
magnetic anisotropy may clarify the reasons for the puzzling
temperature dependencies of $\alpha_i(T)$, $C(T)$, and
$\Gamma_i(T)$.

We acknowledge fruitful discussions with A.~Rosch, I.~Fischer,
and J.A.~Mydosh. This work was supported by the Deutsche
Forschungsgemeinschaft via Sonderforschungsbereich 608.



\begin{thebibliography}{00}
\bibitem{remarkalp}
In zero field there is a small anomaly in
$\alpha_{10\overline{2}}/T$ around
  1~K. At present its origin is unclear, it might be related to the anomaly
  observed in $C/T$ for zero field.
\end{thebibliography}

\end{document}